\documentclass[aps,footinbib,twocolumn,showpacs,amsmath,amssymb,superscriptaddress,prb,groupedaddress]{revtex4}
\usepackage{graphicx,amsmath}
\usepackage{epstopdf}
\usepackage{amssymb}
\usepackage{grffile}
\usepackage[usenames]{color}
\usepackage{indentfirst}
\usepackage{float}
\usepackage{color}
\usepackage{mathrsfs}
\usepackage{dcolumn}
\usepackage{bm}
\usepackage[colorlinks=true,bookmarks=false,citecolor=blue,linkcolor=red,urlcolor=blue]{hyperref}

\begin{document}
\title{Higher-order exceptional lines in a non-Hermitian Jaynes–Cummings triangle}
\author{Hao Chen}
\affiliation{Department of Physics and Chongqing Key Laboratory for Strongly Coupled Physics, Chongqing University, Chongqing 401331, People's Republic of China}
\author{Xiao Qin}
\affiliation{Department of Physics and Chongqing Key Laboratory for Strongly Coupled Physics, Chongqing University, Chongqing 401331, People's Republic of China}
\author{Jian-Jun Dong}
\email{dongjianjun@cqu.edu.cn}
\affiliation{Department of Physics and Chongqing Key Laboratory for Strongly Coupled Physics, Chongqing University, Chongqing 401331, People's Republic of China}
\author{Yu-Yu Zhang}
\email{yuyuzh@cqu.edu.cn}
\affiliation{Department of Physics and Chongqing Key Laboratory for Strongly Coupled Physics, Chongqing University, Chongqing 401331, People's Republic of China}
\author{Zi-Xiang Hu}
\email{zxhu@cqu.edu.cn}
\affiliation{Department of Physics and Chongqing Key Laboratory for Strongly Coupled Physics, Chongqing University, Chongqing 401331, People's Republic of China}

\begin{abstract}
Higher-order exceptional points (EPs) in non-Hermitian systems showcase diverse physical phenomena but require more parameter space freedom or symmetries. It leads to a challenge for the exploration of high-order EP geometries in low-dimensional systems. Here we observe both a third-order exceptional surface and line 
 in a Jaynes-Cummings triangle consisting of three cavities arranged in a ring. A fine-tuning artificial magnetic field dramatically enriches the emergence of the third-order exceptional lines ($3$ELs), which require only three tuning parameters in the presence of chiral symmetry and parity-time (PT) symmetry. Third-order EPs amplify the effect of perturbations through a cube-root response mechanism, displaying a greater sensitivity than second-order EPs. We develop novel fidelity and Loschmidt echo using the associated-state biorthogonal approach, which successfully characterizes EPs and quench dynamics even in PT breaking regime. Our work advances the use of higher-order EPs in quantum technology applications.

\end{abstract}

\date{\today }
\maketitle
\textit{Introduction --} 
Non-Hermitian (NH) systems such as open or dissipative systems are ubiquitous in nature and have generated great interest in gain-loss coupled cavities~\cite{naturephys10,nphys1515,PhysRevLett.114.253601} and photonic platforms~\cite{PhysRevLett.113.053604,PhysRevLett.103.093902}. Non-Hermiticity leads to various counterintuitive phenomena related to EPs that cannot exist in Hermitian systems, such as exceptional nodal topologies~\cite{PhysRevB.104.L201104}, enhanced sensing~\cite{PhysRevLett.117.110802,PhysRevLett.125.240506,PhysRevLett.126.180401}, biorthogonal dynamical quantum phase transition~\cite{PhysRevLett.132.220402,PhysRevLett.122.020501,PhysRevA.98.022129,EPL}, and topological phase transitions~\cite{PhysRevA.98.052116,PhysRevLett.127.090501}. EPs have also been experimentally observed in optical microcavities~\cite{PhysRevLett.103.134101,PNAS2016,PhysRevLett.104.153601} and cold atomic systems~\cite{nphys3842,nphys18}. So far, most of the work has focused on the basic example of second-order EPs ($2$EPs) where two eigenstates coalesce. Great efforts have been made experimentally to explore the $2$EPs, including spontaneous PT symmetry breaking through an EP~\cite{PhysRevLett.126.083604,PhysRevLett.103.093902,PhysRevA.96.043821} and the topological phase with a half-integer winding number~\cite{PhysRevLett.128.160401,PhysRevLett.132.220402}. The more intricate and profound physical phenomena arising from non-Hermiticity are higher-order singularities, emerging from the coalescence of three or more eigenvalues.

Higher-order EPs are of particular interest as they can significantly influence the energy spectrum, stability, and perturbation response. Recent studies have revealed unique advantages of higher-order EPs over $2$EPs notably improved sensitivity~\cite{nature2017,nature2017lan,PhysRevApplied.15.034050}. Higher-order EPs have been reported in optics~\cite{nature2017} and cavity optomechanics systems~\cite{nature2022}. Prior work has primarily focused on isolated higher-order EPs. However, constructing geometries involving higher-order EPs, such as lines composed of higher-order EPs, remains experimentally elusive. Because more degrees of freedom in the Hamiltonian's parameter space or additional symmetries are required. Generally, a $n$th order EP requires $2n-2$ dimensional real parameter space~\cite{RevModPhys.93.015005,PhysRevA.102.032216,PhysRevLett.127.186602}. Theoretical progress has been made that generic NH symmetries can reduce the codimension of 2EPs, enabling symmetry-protected $n-1$ ELs even in $n$-dimensional NH systems ($n\geq1$)~\cite{PhysRevB.99.041202,PhysRevB.99.041406}. Recent experiments have achieved $3$EPs and lines in the symmetry-protected ~\cite{naturenanotechnology} and topological-protected systems~\cite{nctang23}. However, extending these insights to higher-order EP geometries remains an open challenge. 

  \begin{figure}
	\includegraphics[width=8cm]{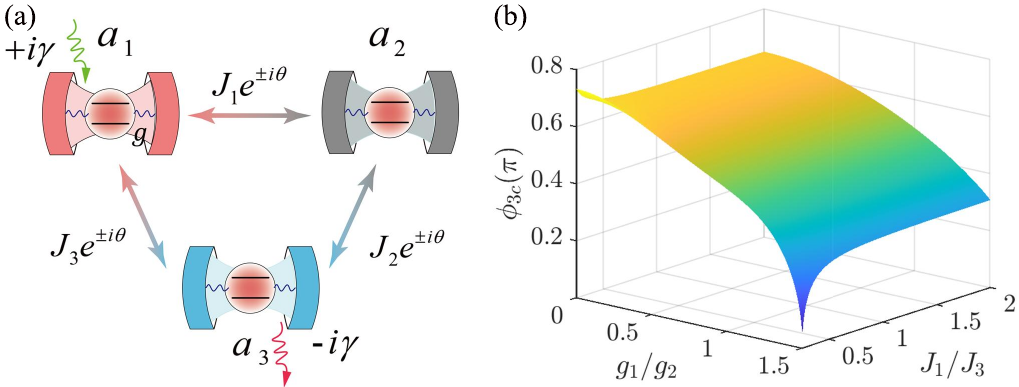}
	\caption{(a) Schematic of the JC triangle with an artificial gauge field $\phi=3\theta$, showing cavity $1$ with gain and cavity $3$ with loss. (b) Third-order exceptional surface $\theta_{3c}$ as a function of $g_1/g_2$ and $J_1/J_3$ for $\omega=1,\Delta=20$.
        }\label{model}
\end{figure}

Attributing to the high tunability of light-atom interactions in cavity QED, we demonstrate how the third-order exceptional surface and lines naturally emerge in a non-Hermitian Jaynes-Cummings (JC) triangle, which is enforced by the PT symmetry. An artificial magnetic field, crucial for inducing chiral symmetry in a coupled-cavity ring, enables exploration of the lines of $3$EPs. Quite remarkably, these singularities are unambiguously identified via biorthogonal fidelity and exhibit enhanced sensitivity to perturbations. The quench dynamics across the EPs are successfully characterized by the biorthogonal Loschmidt echo.

\begin{figure*}[ptb]
\includegraphics[width=16cm]{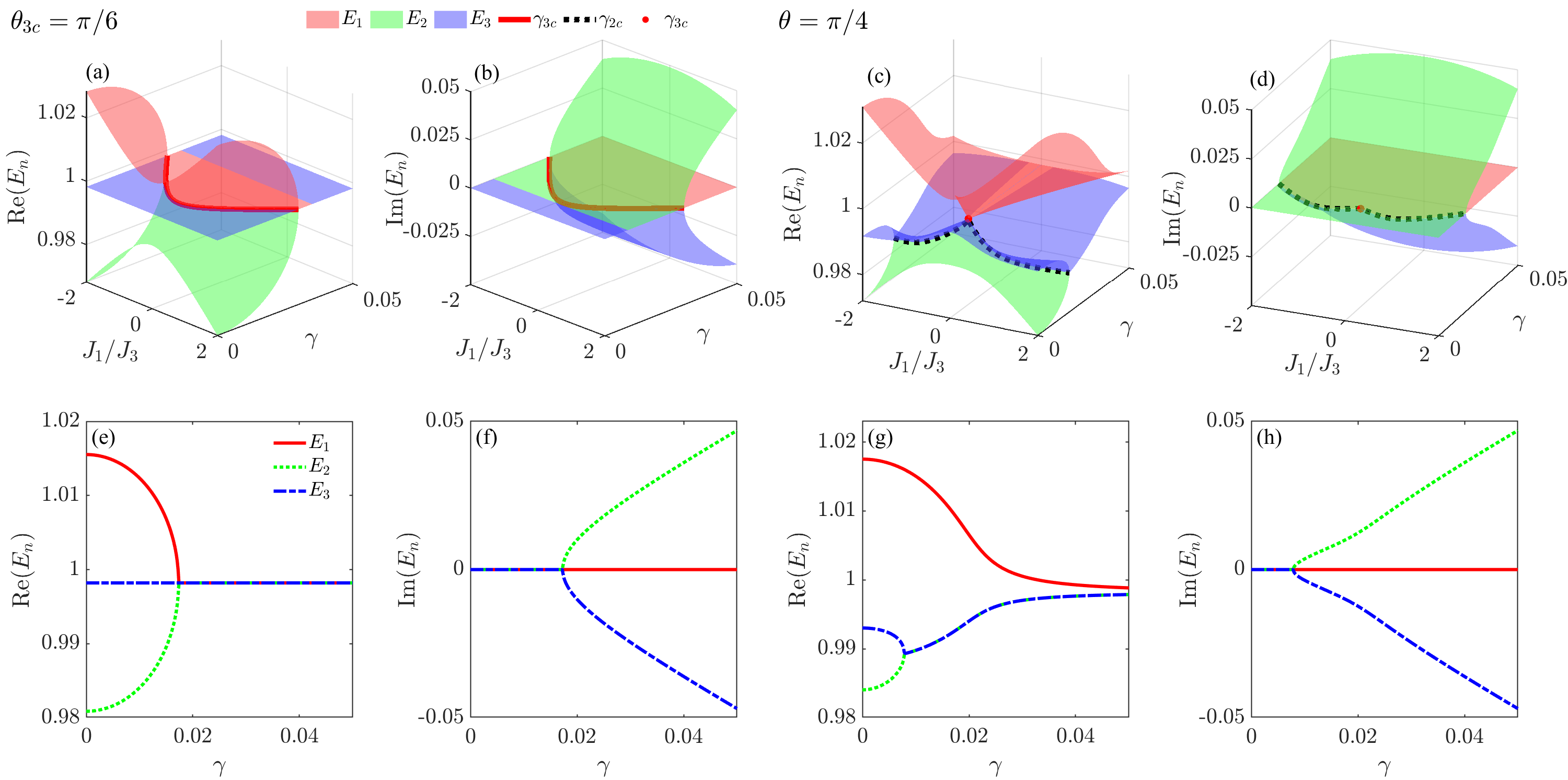}
\caption{The colored surfaces represent the eigenenergies for the real $\mathrm{Re}(E_n)$ (a) and imaginary $\mathrm{Im}(E_n)$ (b) components for $\theta_{3c}=\pi/6$, with respect to gain/loss $\gamma$ and the hopping ratio $J_1/J_3$.The red solid line represents the line of $3$EPs $\gamma_{3c}$. Distinct eigenenergies are shown for $\theta=\pi/4$ in (c) and (d). A blue dot marks a $3$EP, and a black dashed line represents the $2$ELs  $\gamma_{2c}$. For fixed values of $J_1=J_3=0.01$, $\mathrm{Re}(E_n)$ and $\mathrm{Im}(E_n)$ are plotted as functions of $\gamma$, illustrating PT-symmetry breaking at $3$EPs (e)-(f) and $2$EPs (g)-(f). Here, $g_1/\omega=g_2/\omega=0.3$ and $\Delta/\omega=50$. \label{energies}}
\end{figure*}
    
\textit{JC triangle for the third-order exceptional lines --} Fig.~\ref{model}(a) depicts the NH system of the JC triangle, consisting of three connected cavities. The JC Hamiltonian for a cavity interacting with a two-level atom is given by $H_{n}^{JC} = \omega a_n^\dagger a_n + g_n(a_n^\dagger \sigma_n^- + a_n\sigma_n^+) + \Delta\sigma_n^z/2$, where $a_n^\dagger(a_n)$ denotes the creation (annihilation) operator for cavity $n$ with frequency $\omega$, while $\sigma_n^{i}$ are the Pauli matrices representing a two-level atom with an energy gap $\Delta$ and $g$ denotes the strength of the cavity-atom coupling. $H_{n}^{JC}$ possesses $U(1)$ symmetry because the excitation number $N=a^{\dagger}a+\sigma_+\sigma_-$ is conserved.  PT symmetry is achieved in a system of three interconnected cavities: two identical gain or loss cavities and a neutral one in between. The gain medium's lasing increases gain in cavity $1$, while dissipation leads to loss in cavity $3$. The Hamiltonian consists of the local Hamiltonian for each cavity and the photon hopping term
	\begin{align}	H_{JCT}=&\sum_{n=1}^{3}H_{n}^{JC}+\mathrm{i}\gamma a_1^\dagger a_1-\mathrm{i}\gamma a^\dagger_3 a_3\notag\\
		&-J_n(e^{\mathrm{i}\theta}a_n^\dagger a_{n+1}+e^{\mathrm{-i}\theta}a_{n+1}^\dagger a_{n}),
	\end{align}
where $\gamma$ denotes the cavity gain-loss constant, and $J_n$ represents the hopping strength between adjacent cavities $n$ and $n+1$, incorporating a phase $\theta$.
The artificial vector potential $A(r)$ induces complex hopping amplitudes between cavities $n$ and $m$, characterized by the phase $\theta =\int_{r_{n}}^{r_{m}}A(r)dr$. Artificial magnetic flux can be achieved by periodically modulating the photon hopping strength between adjacent cavities~\cite{PhysRevLett.127.063602}.  The magnetic flux across the three cavities is $\phi=3\theta$, maintaining gauge invariance when traversing the closed loop. When $\theta\neq m\pi$ $m\in \mathbb{Z}$ the time-reversal symmetry of hopping processes among three cavities is artificially broken. However, it can be recovered by implementing the chiral transformation $C$ which exchanges the even and odd permutation ($123\leftrightarrow 321$). Consequently, the system has a chiral symmetry, $(CT)H_{JCT}(CT)^{-1}=H_{JCT}$, where the time-reversal operator $T$ satisfies $TiT^{-1}=-i$. Moreover, the system has the PT symmetry, $(PT)H_{JCT}(PT)^{-1}=H_{JCT}$, where the parity operator $P$ obeys $Pa_nP^{-1}=a_{4-n}$.

We perform the Schrieffer-Wolff transformation using $S_n = \exp[g_n/\Delta(a_n^\dagger \sigma_n^{-} - a_n\sigma_n^{+})]$ to eliminate the off-diagonal coupling term in $H_{n}^{JC}$. In the limit $\Delta/\omega\to\infty$, we obtain the transformed Hamiltonian $H_{n}^{JC}=\omega a_n^\dagger a_n+g_n^2/\Delta a_n^\dagger a_n\sigma_n^z+\Delta\sigma_n^z/2$ by omitting higher-order terms. The low-energy Hamiltonian is given by projecting to the subspace of atom $|\downarrow\rangle_n$,
\begin{align}
		H_{JCT}^{\downarrow}=&\sum_n-J_n(e^{\mathrm{i\theta}}a_n^\dagger a_{n+1}+h.c.)\notag\\
		&+\omega_+a_1^\dagger a_1+\omega_2a_2^\dagger a_2+\omega_-a_3^\dagger a_3.
\end{align}
where the renormalized frequency is $\omega_\pm=\omega_{1}\pm\mathrm{i}\gamma$ with $\omega_n=\omega-g_n^2/\Delta$ by setting $g_1=g_3$. The Hamiltonian $H_{JCT}^{\downarrow}$ is expressed bilinearly in terms of bosonic operators and can be diagonalized through the Bogoliubov transformation.  With denotation $\alpha = \{a_1^\dagger, a_2^\dagger, a_3^\dagger\}$, the Hamiltonian is expressed in matrix form as $H^\downarrow_{JCT} = \alpha M \alpha^\dagger$ with a constant, where the matrix $M$ for $J_1=J_2\neq J_3$ is given by
	\begin{align}\label{equation}
		M=\left[
		\begin{matrix}
			\omega_+ & -J_1e^{-\mathrm{i}\theta} & -J_3 e^{\mathrm{i}\theta}\\
			-J_1e^{\mathrm{i}\theta} & \omega_2 & -J_1e^{-\mathrm{i}\theta}\\
			-J_3e^{-\mathrm{i}\theta} & -J_1e^{\mathrm{i}\theta} & \omega_-
		\end{matrix}
		\right].
	\end{align}
 Eigenenergies can be analytically obtained by diagonalizing $M$ via Cardano's formula (see the Supplementary Material (SM)~\cite{supple}). Given $\chi=(-1+\sqrt{3}\mathrm{i})/2$ and $\beta_{\pm}=\sqrt[3]{-q\pm\sqrt{q^2+p^3}}$, the eigenenergies $E_n=\varepsilon_n+(2\omega_1+\omega_2)/3$ are given as 
 \begin{eqnarray}
 \varepsilon_1&=&\beta_{+}+\beta_{-},\notag\\
 \varepsilon_2&=&\chi\beta_{+}+\chi^{*}\beta_{-},\notag\\
 \varepsilon_3&=&\chi^{*}\beta_{+}+\chi\beta_{-},
 \end{eqnarray}
where $p=(-2J_1^2-J_3^2+\gamma^2)/3-(\omega_1-\omega_2)^2/9$, and $q=J_1^2J_3\cos(3\theta)-[\omega_2-\omega_1][(\omega_2-\omega_1)^2+9J_1^2-9J_3^2+9\gamma^2]/27$.

Notably, NH systems support anomalous singularities called EPs, which play a key role in the real-complex spectral transition protected by PT symmetry. The $q^2+p^3$ is the key quantity that determines the EPs. PT-symmetry breaking occurs under the condition $q^2+p^3=0$. When $g_1=g_2$, it yields $2$EPs at the critical value 
	\begin{align}
		\gamma_{2c}=\pm
        \sqrt{3\sqrt[3]{-\left(J_{1}^{2}J_{3}\cos(3\theta)\right)^{2}}+J_{3}^{2}+2J_{1}^{2}}.
	\end{align} 
 Specifically, the $3$EPs appear at the special points where $p=q=0$. It leads to the critical values of the magnetic flux and the gain for the $3$ELs, which are obtained analytically as 
 \begin{eqnarray}
     \theta_{3c}&=&\frac{1}{3}\arccos\left(\frac{(g_2^2-g_1^2)[4(g_2^2-g_1^2)^2/\Delta^2+27J_1^2]}{27\Delta J_1^2J_3}\right), \nonumber \\
     \gamma_{3c}^2&=&2J_1^2+J_3^2+\frac{(g_1^2-g_2^2)^2}{3\Delta^2}.
 \end{eqnarray}
Furthermore, an isolated $3$EP arises while $J_1=J_3=0$ and $g_1=g_2$. Fig.~\ref{model}(b) illustrates the third-order exceptional surface $\theta_{3c}$ by adjusting $g_1/g_2$ and $J_1/J_3$. When $g_1=g_2$, the critical magnetic flux simplifies to $\phi_{3c}=3\theta_{3c}=\pi/2+n\pi$.

\begin{figure}[t] 
		\begin{center}
			\includegraphics[width=8.6cm]{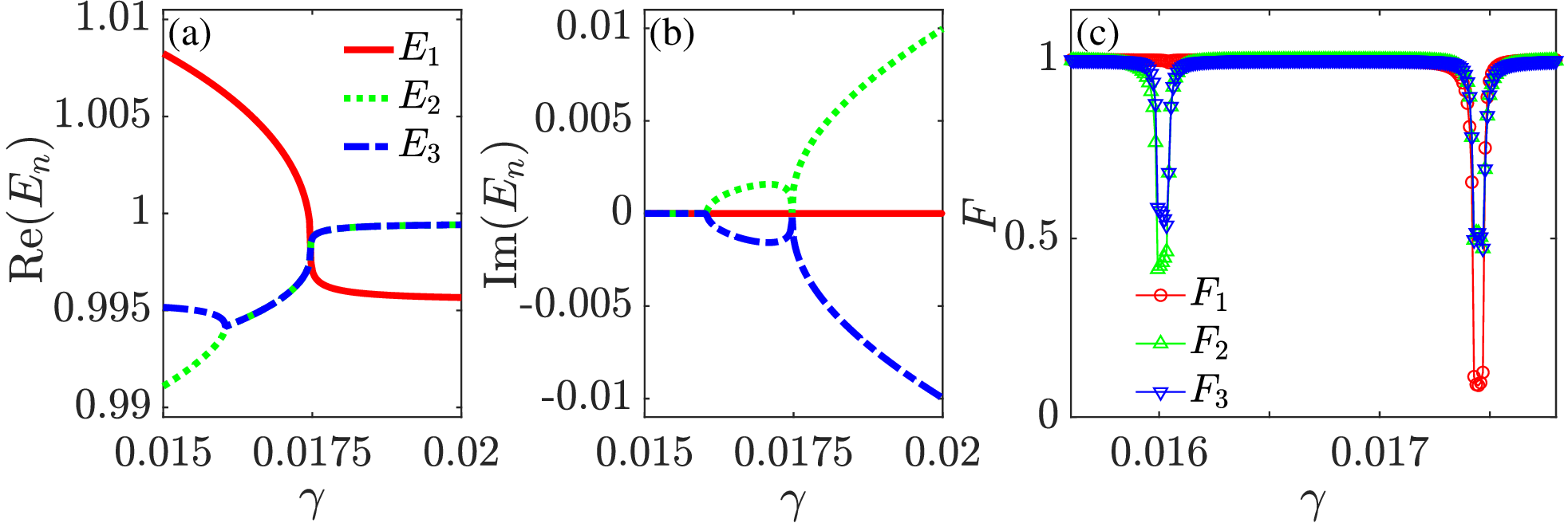}
		\end{center}
		\caption{Real (a) and imaginary (b) parts of $E_n$ as a function of $\gamma$ for the critical magnetic flux $\theta_{3c}$ of a $3$EPs when $g_1/\omega=g_3/\omega=0.1, g_2/\omega=0.3$ and $J_1=J_3=0.01$.(c) Fidelity $F_n$ for three eigenstates between $|\psi_n(\gamma+\epsilon)\rangle$ and $|\psi_n(\gamma)\rangle$ with a small amount $\epsilon=0.00005$. }\label{g1g2}
	\end{figure}	
Fig.~\ref{energies} shows the energy spectrum for different $\theta$ when $g_1=g_2$. At the specific magnetic flux $\theta_{3c}=\pi/6$, the eigenenergy $E_1$ remains real, whereas the other two form complex conjugate pairs above the critical value $\gamma_{3c}$, satisfying $\mathrm{Im}[E_2]=-\mathrm{Im}[E_3]$. The line of 3EPs $\gamma_{3c}$ is located on the curve defined by $q^2+p^3=0$ and $q=0$, where the real parts of three eigenvalues become equal, $\mathrm{Re}[E_1]=\mathrm{Re}[E_2]=\mathrm{Re}[E_3]$, as depicted in Fig.~\ref{energies}(a) and (e). It shows three-state coalescence for any $J_1/J_3$. For $\gamma>\gamma_{3c}$ and $q^2+p^3>0$, Fig.~\ref{energies}(b) and (f) illustrate that two energies form a complex conjugate pair, where $\mathrm{Im}[E_2]=-\mathrm{Im}[E_3]$. Otherwise, for $\theta\neq\theta_{3c}$, the line of $2$EPs arises from condition $q^2+p^3=0$ with $q\neq 0$, as illustrated in Fig.~\ref{energies}(c) and (d). An isolated 3EP resides at the intersection of two 2ELs. Along the 2ELs $\gamma_{2c}$, the real components of two eigenvalues merge, whereas $E_1$ remains distinctly real in Fig.~\ref{energies}(g)-(h).

The $3$ELs can be fine-tuned by modifying the atom-cavity coupling in each cavity. The $3$EP bifurcates into an additional $2$EP by tuning the atom-cavity coupling $g_1/g_2=1/3$ in Fig.~\ref{g1g2}(a)(b). Because of different coupling strengths, the eigenstates of the gain and loss cavities coalesce and generate a $2$EP, which is separated from the neutral cavity $2$ in the eigenenergy. This features an EP formation with tunable cavity-dependent coupling strengths.

\begin{figure}[ptb] 
		\begin{center}
			\includegraphics[width=8.6cm]{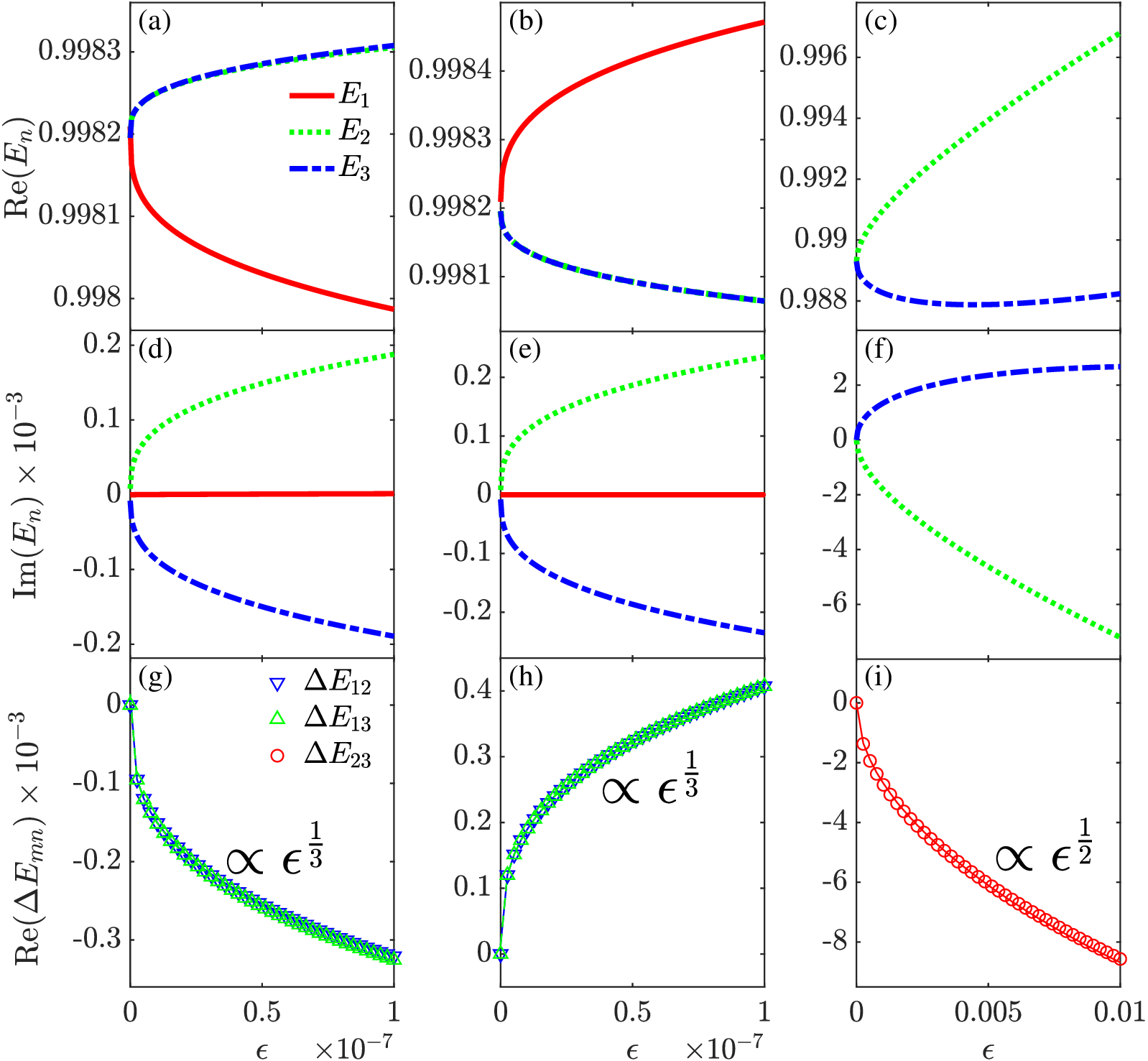}
		\end{center}
		\caption{(a)-(f)Real and imaginary parts of $E_{n}$, and the energy difference $\mathrm{Re}(\Delta E_{12})$, $\mathrm{Re}(\Delta E_{13})$ at the $3$EP ($\theta_{3c}=\pi/6$) for the perturbation $\epsilon$ on the gain cavity $1$ (left column) and the neutral one $2$ (middle column), respectively.  Right column shows $E_{n}$ and $\mathrm{Re}(\Delta E_{23})$ at the $2$EP ( $\theta=\pi/4$) for the perturbing on the gain cavity.
        }\label{sensitivity}
	\end{figure}
    
\textit{Enhanced sensitivity along the lines of $3$EPs-} 
To measure the enhanced sensitivity at the $3$EPs, we introduce a disturbance to the gain cavity ($\epsilon_1=\epsilon$ and $\epsilon_{2,3}=0$). The eigenenergies $E_{n}$ equation are given by $ E_n^{3}-\left(a+\epsilon\right)E_n^{2}+\left(b+c\epsilon\right)E_n+d\epsilon+u=0$ (see the SM~\cite{supple}).
 Comparing to the  unperturbed eigenenergy $E_0$ at $\gamma_{3c}$, the energy difference $\Delta E_n = E_n - E_0$ constitutes a small correction. We employ a Newton-Puiseux expansion to express this difference as $\Delta E_n = c_1 \epsilon^{1/3} + c_2 \epsilon^{2/3} + \ldots$ (see SM~\cite{supple})
 \begin{equation}
 \Delta E_n \sim e^{\mathrm{i}(2n+1)\pi/3}\eta^{1/3}\epsilon^{1/3}-\frac{v}{3\eta^{1/3}}e^{\mathrm{-i}(2n+1)\pi/3}\epsilon^{2/3},
 \end{equation}
 where $\eta=-E_0^2+(\omega_{-}+\omega_0)E_0+J^2-\omega_0\omega_{-}$ and $v=\omega_{-}+\omega_0-2E_0$. The energy difference $\mathrm{Re}(\Delta E_{12(13)})$ between $E_{1}$ and $E_{2(3)}$ follows the cube root of $\epsilon$, while the splitting between $\mathrm{Re}(E_{2})$ and $\mathrm{Re}(E_{3})$ is very small and is on the order of $\epsilon^{2/3}$.  $\mathrm{Re}(\Delta E_{12(13)})$ are analytically expressed as 
   \begin{eqnarray}\label{cuberoot}
    \mathrm{Re}(\Delta E_{12(13)}) &\sim& \frac{3}{2}\eta^{1/3}\epsilon^{1/3}\pm\frac{\sqrt{3}\gamma_{3c}}{6}\eta^{-1/3}\epsilon^{2/3}.
    \end{eqnarray}
This cube-root scaling indicates an increased sensitivity to perturbations.
Fig.~\ref{sensitivity} (a) and (d) show the real and imaginary components of the eigenenergies at the 3EP with $\theta_{3c}=\pi/6$ for the perturbing gain cavity.  When $\epsilon$ appears on the neutral cavity, a distinct change of eigenenergies is observed in Fig.~\ref{sensitivity} (b) (e). Both of the scaling behaviors of the energy splitting show clear cube-root at the 3EP in Fig.~\ref{sensitivity} (g)-(h), which fitting well with the analytical results in Eq. (\ref{cuberoot}). For the $2$EPs, the energy splitting $\Delta E_{2(3)}=E_{2(3)}-E_0$ can be expanded in terms of $\epsilon$ as $\Delta E_{2(3)} \sim c_1 \epsilon^{1/2} + c_2 \epsilon$. It leads to the energy splitting as 
   \begin{eqnarray}
   \mathrm{Re}(\Delta E_{23})&\sim&\mathrm{Re}(-2\sqrt{\alpha})\epsilon^{1/2},
    \end{eqnarray}
where $\alpha=-\sqrt{9(E_0-\omega_{+})^2-5(\omega_0+\omega_{-}-2E_0)^2}/5-3(E_0-\omega_{+})/5$.
Figs.~\ref{sensitivity}(c)(f) shows eigenenergies varing with $\epsilon$ at the $2$EP for the perturbing gain cavity. The energy difference $\Delta E_{23}$ exhibits the square-root behavior in Fig.~\ref{sensitivity}(i). The above comparison reveals increased perturbation sensitivity of $3$EPs.
 

\textit{Associated-state biorthogonal Fidelity and Loschmidt Echo--} 
Fidelity is used as an estimation of the similarity of two
quantum states, exhibiting a quick drop at critical points. However, the fidelity loses its conventional meaning in non-Hermitian system, as the standard inner product of quantum mechanics leads to unphysical behavior. For a non-Hermitian Hamiltonian $H\neq H^{\dagger}$, the eigenvalues equations of $H$ and $H^{\dagger}$ are given by $H|\psi_{n}\rangle=E_n|\psi_{n}\rangle$ and $H^{\dagger}|\widetilde{\psi}_{n}\rangle=E_n^{*}|\widetilde{\psi}_{n}\rangle$, where  $|\psi_{n}\rangle$  and $\langle\widetilde{\psi}_{n}|$ are the
right and left eigenstates of $H$ that satisfy the biorthonormal
relation $\langle\widetilde{\psi}_{m}|\psi_{n}\rangle=\delta_{mn}$ and the completeness relation $\sum_{n}|\widetilde{\psi}_{n}\rangle\langle\psi_{n}|=1$. To generalize
the ﬁdelity for non-Hermitian systems, we propose the biorthogonal Fidelity in the framework of the associated-state biorthogonal method 
\begin{equation}
F(\gamma)=\frac{\langle\widetilde{\psi}(\gamma)|\psi(\gamma+\epsilon)\rangle\langle\widetilde{\psi}(\gamma+\epsilon)|\psi(\gamma)}{\langle\widetilde{\psi}(\gamma+\epsilon)|\psi(\gamma+\epsilon)\rangle\langle\widetilde{\psi}(\gamma)|\psi(\gamma)\rangle}.
\end{equation}
For the perturbed state $|\psi(\gamma+\epsilon)\rangle$ with a small perturbation $\epsilon$, the associated state $|\widetilde{\psi}(\gamma+\epsilon)\rangle$ is defined according to the following relation ~\cite{PhysRevLett.132.220402}
\begin{equation}
|\psi(\gamma+\epsilon)\rangle=\sum_na_n|\psi_n(\gamma)\rangle \leftrightarrow |\widetilde{\psi}(\gamma+\epsilon)\rangle=\sum_na_n|\widetilde{\psi}_n(\gamma)\rangle.  
\end{equation}
This representation indicates $F(\gamma) \in [0,1]$, thus eliminating the unphysical fidelity value at EPs reported in prior studies~\cite{PhysRevResearch.3.013015} (see the SM~\cite{supple}). Remarkably, all three eigenstates show a significant decline at the $3$EP $\gamma_{3c}$, while two exhibit a marked decrease at $\gamma_{2c}$ in Fig.~\ref{g1g2}(c).

Similarily to the fidelity, the Loschmidt echo is used to quantify the sensitivity of quantum evolution to perturbations.
For a quantum quench dynamics, the Hamiltonian changes abruptly from $H^{i}$ at $t=0^{-}$ to $H^{f}$ at $t=0^{+}$. The initial eigenstate $|\psi(0)\rangle=|\psi_{n}^{i}\rangle$ of $H^{i}$ evolves under the Hamiltonian $H^{f}$ as $|\psi(t)\rangle = e^{-\mathrm{i} H^{f}t}|\psi(0)\rangle$, while the direct generalization $|\widetilde{\psi}(t)\rangle=e^{-\mathrm{i}H^{f\dagger}t}|\widetilde{\psi}(0)\rangle$ leads to unreasonable complex probabilities in the PT breaking regime~\cite{EPL,PhysRevLett.132.220402}. For the time-evolved state $|\psi(t)\rangle$, we define its associated state as \cite{PhysRevLett.132.220402}
\begin{equation}
 |\widetilde{\psi}(t)\rangle=\sum_{n}d_{n}|\widetilde{\psi}_{n}^{i}\rangle,d_{n}=\langle\widetilde{\psi}_{n}^{i}|\psi(t)\rangle.   
\end{equation}
The overlap between $|\psi(0)\rangle$ and $|\psi(t)\rangle$ is characterized by the associated-state biorthogonal Loschmidt echo
\begin{equation}
\mathcal{L}\left( t\right)  =\frac{\langle\widetilde{\psi}(0)|\psi%
(t)\rangle\langle\widetilde{\psi}(t)|\psi(0)\rangle}{\langle
\widetilde{\psi}(t)|\psi(t)\rangle\langle\widetilde{\psi}(0)
|\psi(0)\rangle}.
\end{equation}
We obtain $\mathcal{L}_{n}\left(  t\right)\in [0,1]$ for three initial eigenstate $|\psi_n^i\rangle$ as in the Hermitian case.

 \begin{figure}[pt]
			\includegraphics[width=8cm]{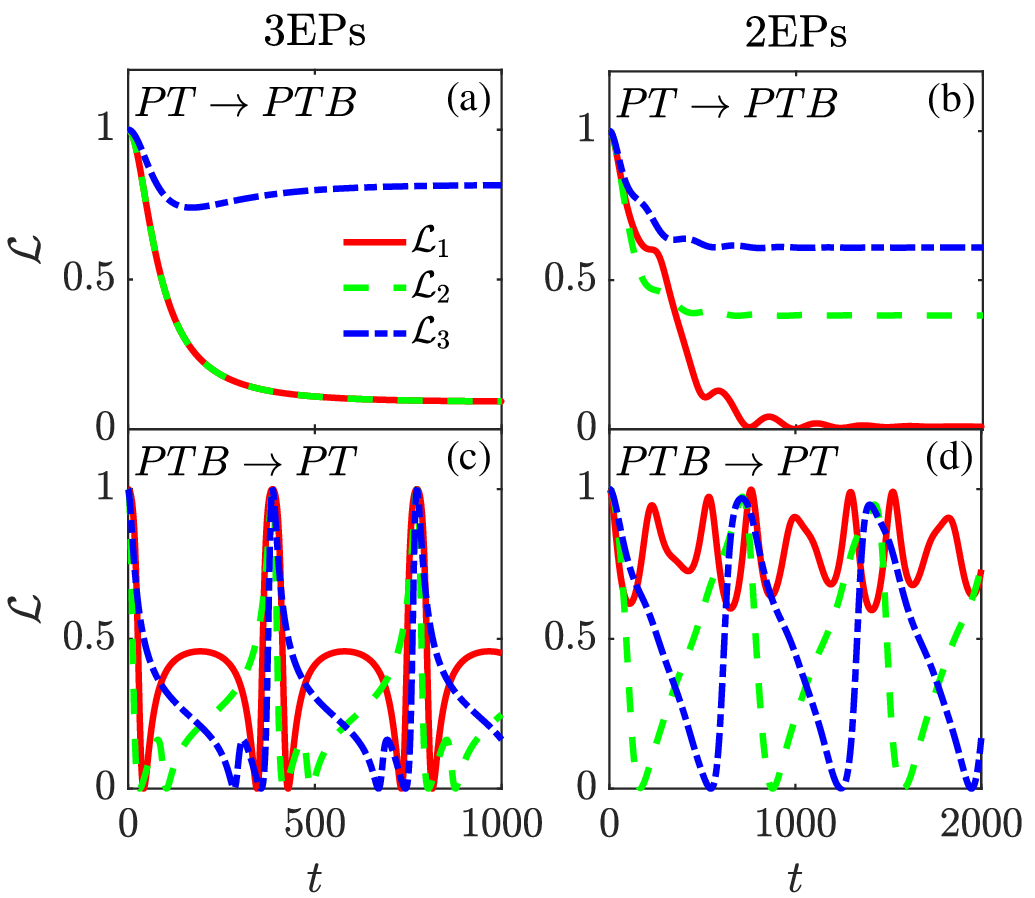}
		\caption{Loschmidt echo of quenching processes across the $3$EP ($\theta=\pi/6$) (a) from $\gamma_i=0.006$ to $\gamma_f=0.018$ and its reversal dynamics (c). The quench dynamics cross the $2$EPs ($\theta=\pi/4$) (b) from $\gamma_i=0.001$ to $\gamma_f=0.01$ and its reversal process (d). The initial states are chosen three eigenstates.}
		\label{LE}
	\end{figure}

In the PT symmetric regime of the prequench Hamiltonian $H^{i}$, two identical curves, $\mathcal{L}_{1}(t)$ and $\mathcal{L}_{2}(t)$, arise when $\gamma$ quench across the $3$EP, while distinct curves $\mathcal{L}_{n}(t)$ are observed when the $\gamma$ quench across the $2$EP in Fig. \ref{LE}(a)(b). The identical curves likely stem from the symmetry present in the eigenenergy spectrum. The significance of the postquench Hamiltonian $H^{f}$ arises from the inclusion of the term $e^{-\mathrm{i} E_{n}^{f}t}$ in $d_{n}$. When $H^{f}$ resides in the PT symmetric regime, the $\mathcal{L}_{n}(t)$ exhibits periodic oscillations, while it tends
to a steady state in the PT-broken regime due to the complex eigenenergy $E_{n}^{f}$.
 
\textit{Conclusion--} 
We have presented an unprecedented example of a novel non-Hermitian light-atom interaction system exhibiting high-order EP geometry within a one-dimensional JC triangle. The JC triangle, with its artificial magnetic field and the interplay of PT symmetry and chiral symmetry, reveals a complexity of EP-related phenomena. Unique singularities at lines of $3$EPs exhibit greater sensitivity over $2$ELs. Novel associated-state biorthogonal fidelity and Loschmidt echo are proposed  to characterize the singularities and distinct dynamics across the $3$EP even in the PT-broken regime. The realization of high-order  exceptional lines with improved sensitivity suggests promising future applications in quantum technologies and ultra-sensitive sensors related to light-matter interactions.

\textit{Acknowledgments--} This work was supported by NSFC under Grants Nos. 12075040, 12147102, 12474140, 11974064, 12347101, Chongqing Natural Science Foundation Grant No. cstc2020jcyj-msxmX0890 and the Fundamental Research Funds for the Central Universities Grant No. 2024CDJXY022.

\bibliography{refs}{}

\end{document}